\def\G1915{GRS $1915$+$105$}
\def\X1550{XTE J$1550$--$564$}
\def\J1655{GRO J$1655$--$40$}

\def\etal{et al. }

\documentclass[]{aastex}
\usepackage{emulateapj5}
\usepackage{epsfig}
\usepackage{epsf}
\usepackage{color}
\definecolor{red}{rgb}{0.7,0,0}
\definecolor{blue}{rgb}{0,0,0.7}

\shorttitle{LFQPO Spectra of \G1915}
\shortauthors{Rodriguez \etal}
\begin{document}
\title{Spectral Properties of Low Frequency Quasi-Periodic
Oscillations in GRS~1915+105}

\author{J. Rodriguez\altaffilmark{1,2}, S. Corbel\altaffilmark{3,1}, D.C Hannikainen\altaffilmark{4}, T. Belloni\altaffilmark{5}, A. Paizis\altaffilmark{2}, O. Vilhu\altaffilmark{4}}

\altaffiltext{1}{DSM/DAPNIA/Service d'Astrophysique (CNRS FRE 2591), CEA Saclay, 91191 Gif sur Yvette, France}
\altaffiltext{2}{ISDC, Chemin d'Ecogia, 16, 1290 Versoix, Switzerland}
\altaffiltext{3}{Universit\'e Paris 7 Denis Diderot, 2 Place Jussieu, 75005 Paris, France}
\altaffiltext{4}{Observatory, PO Box 14, FIN-00014, University of Helsinki, Finland}
\altaffiltext{5}{INAF - Osservatorio Astronomico di Brera via E. Bianchi 46, 23807 Merate (LC), Italy}

\begin{abstract}
We report on the timing analysis of {\emph{RXTE}} observations
of the Galactic micro-quasar \G1915 performed in 2003. 
Out of a total of six times $\sim 20$~ks, we focus here 
only on the three observations during which  \G1915 is found in 
a steady C-state (referred to as class $\chi$) resulting 
in a total of $\sim 50$~ks. During these observations, we detect 
low frequency quasi-periodic oscillations 
with high ($\sim 14\%$) rms amplitude in the 2--40
keV energy range. Contrary to what is usually observed in GRS~1915+105, in most 
of our observations the QPO frequency present no correlation with the {\emph{RXTE/PCA}}
 count rate, nor with the {\emph{RXTE/ASM}} count rate. We present, for the first time, 
high resolution (22 spectral channels) 
2-40 keV spectral fits of the energy dependence of the QPO amplitude (``QPO spectra''). 
The QPO spectra are well modeled with a cut-off power law except on one occasion 
where a single power law gives a satisfactory fit (with no cut-off at least up to $\sim 40$ keV). 
The cut-off energy evolves significantly from one observation to the other, from a value 
of $\sim21.8$ keV  to $\sim30$ keV in  the other observations where it is detected.  
We discuss the possible origin of this behavior and suggest that the compact jet 
detected in the radio contributes to the hard X-ray ($\geq 20$ keV) mostly through synchrotron 
emission, whereas the X-ray emitted below 20 keV would originate through inverse Compton scattering.
The dependence of the QPO amplitude on the energy can be understood if the modulation of
the X-ray flux is contained in the Comptonized photons and not in the synchrotron ones. 
\end{abstract}


\section{Introduction}
\G1915 was discovered by the {\emph{WATCH}} instrument on-board 
{\emph{GRANAT}}
in 1992 (Castro-Tirado et al. 1992). It is the first 
Galactic source observed to have apparent superluminal 
motion in radio (Mirabel \& Rodr{\'\i}guez, 1994), 
corresponding to the ejection of plasma at a speed of 
$\sim 92-98\%$ of the speed of light. It distance is 
estimated to $9\pm3$ kpc (Chapuis \& Corbel 2004), and the mass of the 
compact object
in \G1915 is estimated to be 14.0 $\pm 4.4$ M$_{\odot}$ 
(Harlaftis \& Greiner 2004).\\
\indent Systematic monitoring in the X-rays (mainly with 
the {\emph{Rossi X-ray Timing Explorer, RXTE}}), revealed a rich
pattern of variability on all time scales. 
\G1915 is a source of Low and High Frequency QPOs (LFQPO, 
HFQPO, Morgan, Remillard \& Greiner 1997), whose properties 
(frequency, rms amplitude) are tightly correlated to the 
spectral parameters (Morgan et al. 1997, 
Muno et al. 1999, Markwardt et al.
1999, Rodriguez et al. 2002a,b, Vignarca et al. 2002).  
When analyzing data of black hole binaries, the frequency of LFQPOs  
have been shown to be best correlated with 
the slope of the high energy tail of the energy spectra 
(Vignarca et al. 2002). It should be noted that  the 
LFQPO frequency is usually correlated with the soft X-ray flux, thought
to originate from the accretion disk.\\
\indent Belloni et al. (2000, hereafter B00) analyzing 163 
{\emph{RXTE}} observations, have shown that, 
though complex, the behavior of \G1915 could be understood
as spectral transitions between 3 basic states A, B, C. 
They identified 12 recurrent classes of variability 
on a timescale of $\sim$ 3000 s. \G1915 spends most of the time
in the so-called $\chi$ class of variability that corresponds to a 
steady state in the X-rays, lying in a rather hard part 
of the color-color diagram (state C or hard 
state). 
Based on the X-ray (spectral end temporal) and radio 
properties of \G1915, 4 subclasses ($\chi_1$, $\chi_2$, 
$\chi_3$, $\chi_4$) can be distinguished. 
Two of them have a high level of radio emission with a flat 
spectrum, LFQPOs, and a high energy tail (B00, 
Trudolyubov 2001, Muno et al. 2001, Klein-Wolt et al. 
2002).\\ 
\indent We monitored \G1915 with the {\emph{INTErnational 
Gamma-Ray Astrophysics Laboratory (INTEGRAL)}} during its 
first AO, for a total of 6$\times$100 ks (Hannikainen et 
al. 2003, 2004), and obtained 120 ks divided in 6 sequences of 
simultaneous observations with {\emph{RXTE}}. 
One of our {\emph{RXTE}} observations was planned during an {\emph{INTEGRAL}}
target of opportunity on \G1915, and allowed a wide band
simultaneous spectral and temporal coverage to be
performed (Fuchs et al. 2003, hereafter F03).  The global analysis of the 
whole campaign is dedicated to a future publication. Here we 
focus on the timing analysis of the 3 steady
C-state {\emph{RXTE}} observations. The data reduction methods are 
described  in Sec. 2, while the results are presented in Sec. 3
and discussed in Sec. 4

\section{Observations and Data Reduction} 
The log of the observations analyzed in this paper can 
be found in Table \ref{tab:log}.  Each observation 
covers several satellite orbits.
The analysis was first performed on each single revolution,
and when no noticeable (spectral or temporal) evolution 
was found, the different orbits were further averaged to
increase the statistics.

\begin{figure}
\plotone{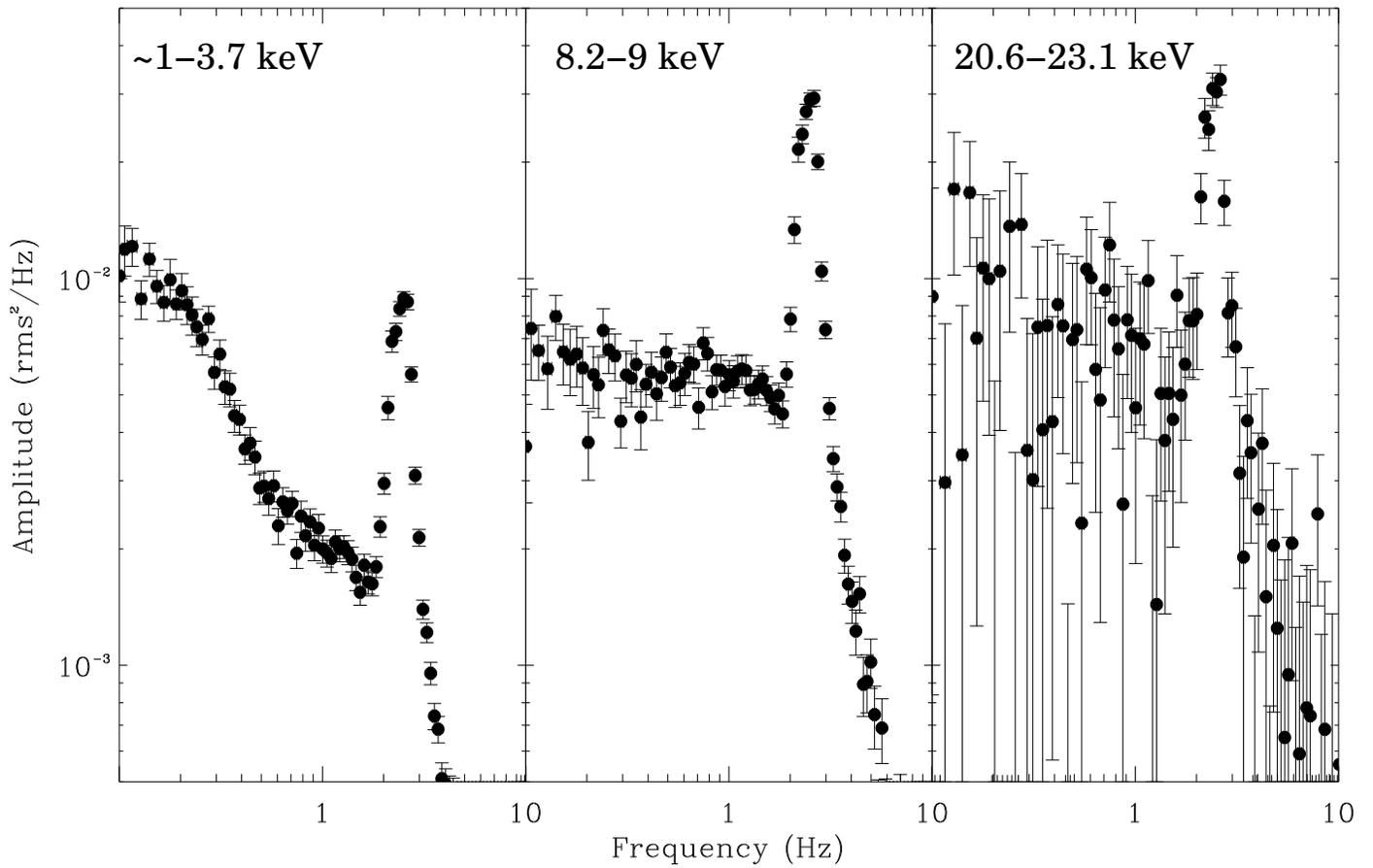}
\caption{Example of PDS's extracted in three energy bands (1--3.7 keV, 8.2--9 keV, and 20.6-23.1 keV), 
as  described in the text. These PDS's are from Obs. 1. The $2.498$ Hz LFQPO is obvious in each 
panel. The same vertical scale is used for each PDS and allows for a direct comparison of the 
source behavior in those energy bands.}
\label{fig:pds}
\end{figure}

We extracted light curves  from the 
{\emph{Proportional Counter Array (PCA)}} following the standard 
way described in the Cook Book and ABC of {\emph{RXTE}}, with the
{\emph{LHEASOFT V5.3}} package. 
Good time intervals (gti) are defined as follows:
satellite elevation over the Earth limb $>10^{\circ}$,
offset pointing $<0.02^{\circ}$, and PCU 0 and 2 turned
on. Light curves were extracted from Binned and Event data. 
We first accumulated a broad band 2--40~keV (absolute channels 0-94, 
epoch 5) light curve with the highest time resolution allowed by 
the (Binned) data format ($\sim 4$~ms). We then extracted light curves 
in small energy bins, with the highest spectral resolution 
allowed by the (Binned) data (16 energy bins from 2 to 14.8 keV), and over
7 additional spectral bins for the Event data (from 15 to $\sim$40~keV). 
Power density spectra (PDS) were produced using 
{\itshape{Powspec V1.0}}, and corrected for white noise.
In the case of the 2--40 keV light curves, PDS were produced on 
interval length of 64~s between 
15.625~mHz and 64~Hz. All intervals were averaged together.
The energy dependent PDSs were produced on interval of 160~s length,
between 6.25~mHz and 12.8~Hz. Fig. \ref{fig:pds} shows, as an example,
the PDS's extracted in three energy bands, for Obs.1
We extracted background light curves in all these energy ranges, 
 and used their count rate to obtain the true 
rms amplitude following Berger and van der Klis (1994). In addition, 
to check for short term evolution of the QPO frequency, we produced dynamical 
power spectra  with $\sim 16$~s resolution between absolute channels 0-35 
($\sim2-14.8$~keV).  \\

\section{Results}
\begin{figure}
\plotone{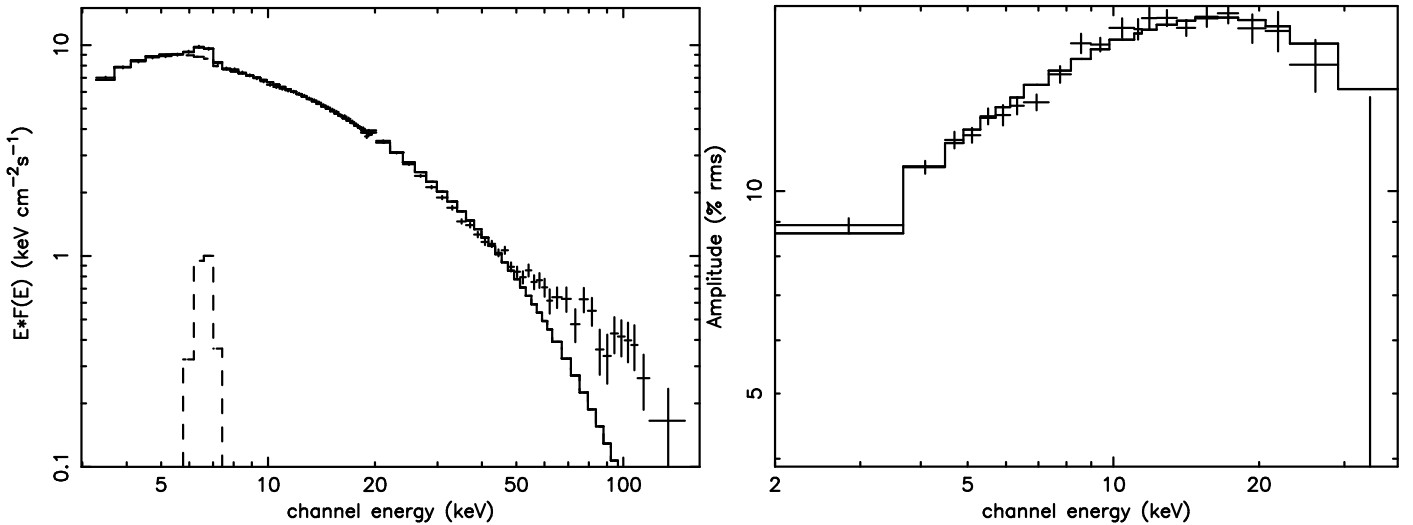}
\caption{{\it Left panel}: PCA+HEXTE 3-150 keV $\nu-f_\nu$ spectra from Obs. 1. A model consisting 
of an absorbed power law with a high energy cut-off of about 25 keV (plus a Gaussian at 6.4 keV) is 
superimposed. The deviation to this  model at high energy is obvious here. This likely reflects 
the need of an additional component to account for the high energy emission from GRS 1915+105.
{\it Right panel}: Energy dependence of the QPO amplitude observed during the same observation.
The best fit model a power law with a high energy cut-off of $\sim 22$~keV is superimposed.}
\label{fig:spec}
\end{figure}

The preliminary spectral  analysis (in a multi-wavelength context) 
of the first observation is presented in F03. 
The 8 other sequences presented here show similar steady light 
curves (see Hannikainen et al. 2004, hereafter H04, for details on the 
{\emph{INTEGRAL/RXTE}} campaign). The {\emph{RXTE/ASM}} light curve, showing
the dates of our pointed {\emph{INTEGRAL/RXTE}} observations is represented in 
Fig. {\ref{fig:asm}}. While the long term evolution shows a slow decay, a double flare
occurs between Obs.~1 and Obs.~2-7. This X-ray flare is associated with a radio flare 
(F03), probably indicative of a discrete ejection.
We identify the class of variability of our observations as class $\chi$ of B00.  
The high level of radio 
emission detected during each of these observations (F03, H04) allows us to further 
classify the observations as class $\chi_1-\chi_3$ also known as
the radio loud hard state (Muno et al. 2001) or type II hard state (Trudolyubov 2001).
It should be noted, however, such a long term decay with the source mostly in class $\chi$ 
is rather peculiar, and had never been seen previously. 
A preliminary spectral analysis (F03; H04) shows that the common model
of black hole X-ray binaries, i.e. a multicolor disk black body  and 
a power-law, represents the data well. As mentioned for such classes 
(Muno et al. 2001), however, the disk temperature returned from the fit 
is too high (3-4 keV), and the inner radius far too small. Alternative models 
of broken power-law or broken power-law plus 
disk component fit the data well and lead to parameters closer to what 
is seen in other systems (H04). We also successfully fitted the {\emph{RXTE/PCA}} 3-25 keV 
spectra with a cut off power-law with a high energy cut-off of about 20-25 keV 
(Rodriguez et al. 2004, Fig \ref{fig:spec}). When adding higher energy spectra, 
such as those extracted with {\emph{RXTE/HEXTE}}, a large deviation to the spectrum is 
seen at high energy, indicating the need of an additional spectral component to the model, 
e.g. an extra power law (Zdziarski et al. 2001; Hannikainen et al. in prep.). 
This is illustrated in 
Fig. \ref{fig:spec} (left panel) with the particular example of Obs. 1\footnote{Note 
that the details of the RXTE and INTEGRAL spectral analysis will 
be given in a forthcoming paper, dedicated to the spectral analysis of the whole campaign.
However, the RXTE (PCA, and HEXTE) spectra have been extracted in the same way as in, e.g., 
Rodriguez et al. 2003.}. Furthermore, Rodriguez et al. (2004) have shown that the 20-400 keV
combined {\emph{RXTE/HEXTE}} and {\emph{INTEGRAL/IBIS}} and {\emph{SPI}} spectra could be 
fitted with a power law of photon index $\sim 3.5$. Note that similar results were found from the 
OSSE spectral analysis of Zdziarski et al. (2001).
 
For all sequences, the 2--40~keV PDS were fitted between $\sim$15~mHz
and 10~Hz  with a sum of 2 or 3 Lorentzians (depending on the energy range), 
to account for the wide band variability (Belloni, Psaltis \& van der Klis 2002). 
A strong LFQPO is detected in all the 
PDSs, and is modeled with an additional Lorentzian (harmonics are also detected, 
especially during intervals with the longest exposures).  A first analysis of Obs. 8
showed a rather broad QPO, with parameters poorly constrained.
As the dynamical power spectrum showed two distinct features, we separated this 
observation in sub-intervals 
and averaged those showing the QPO at the same frequency. This resulted in two 
distinct sets of data, for which we identified two different QPOs. 
The LFQPO parameters are reported in Table \ref{tab:qpos}. \\
\indent At first glance, there is apparently no obvious correlation between the 
QPO frequency and the PCA 2--60 keV count rate (Tables \ref{tab:log} \& 
\ref{tab:qpos}). To further verify this, we fitted each of the $\sim2-15$~keV  16~s PDS 
used to construct our dynamical power spectra with a Lorentzian around the average QPO frequency 
(Table \ref{tab:qpos}), and could therefore obtain the variations of the QPO frequency
with a time resolution of 16~s. No correlation is found between the QPO frequency and 
the PCA $2-15$~keV count rate from Obs. 1 through 7, whereas we do find 
a correlation in Obs. 8.\\
\indent We further averaged sequences showing the QPO at a similar
frequency (Obs. 7, showing the QPO at 1.06 Hz is averaged with Obs. 2
and Obs. 4, whereas  Obs. 3, 5 and 6 are averaged together), and produced PDS in 
the 22 energy bins described in Sec. 2. 
These energy dependent PDSs were fitted 
between 6.25~mHz and 10~Hz. The energy dependences of the amplitude
of the four distinct features are reported in Fig. \ref{fig:qpospec}.\\
\indent A clear difference in the shape of the energy dependence of the amplitude 
of the QPO appears in Fig. \ref{fig:qpospec}. A clear turn-over in the amplitude 
vs. energy relation is visible for the $\sim2.48$~Hz QPO 
detected on MJD 52731, and another one is visible for the $\sim1.09$~Hz QPO 
from the observation of MJDs 52738-52739, although it is not as clear as for the first 
QPO. For the three other features we do not see any clear turn-over 
 (Fig. \ref{fig:qpospec}), although a flattening is obvious at  energies above 10 keV.
 This may suggest that the turn-over energy evolves from one observation to the other, 
and is above the upper energy limit of our QPO ''spectra''. To further test this hypothesis,
we fitted the QPO spectra in {\emph{XSPEC V11.3.0}}. For all QPOs but the 1.04 Hz one, the
spectra are well fitted by a cut-off power-law ({\rm{cutoffpl}} in {\emph{XSPEC}}). The fit 
parameters are reported in Table \ref{tab:fit}, while the right panel of Fig. 
\ref{fig:spec} shows the QPO spectrum of Obs. 1 with the best fit model superimposed. 
It should be added here that a single power law 
gives a rather good representation of the 1.878~Hz QPO detected in Obs. 8, with a 
reduced chi square of 1.89 (20 dof). A cut-off power law model improve the fit (Table 
\ref{tab:fit}), although the cut-off energy is poorly constrained (4.7-$\sigma$ significance 
on this parameter). It is interesting to note that the break energy seems anti-correlated with 
the QPO frequency, i.e. the lowest break energy is observed for the highest QPO frequency 
(Table \ref{tab:fit}). Caution has to be expressed, however, since the statistical uncertainties
on the break energies does not allow us to draw firm conclusion.


\section{Discussion}
\begin{figure}
\plotone{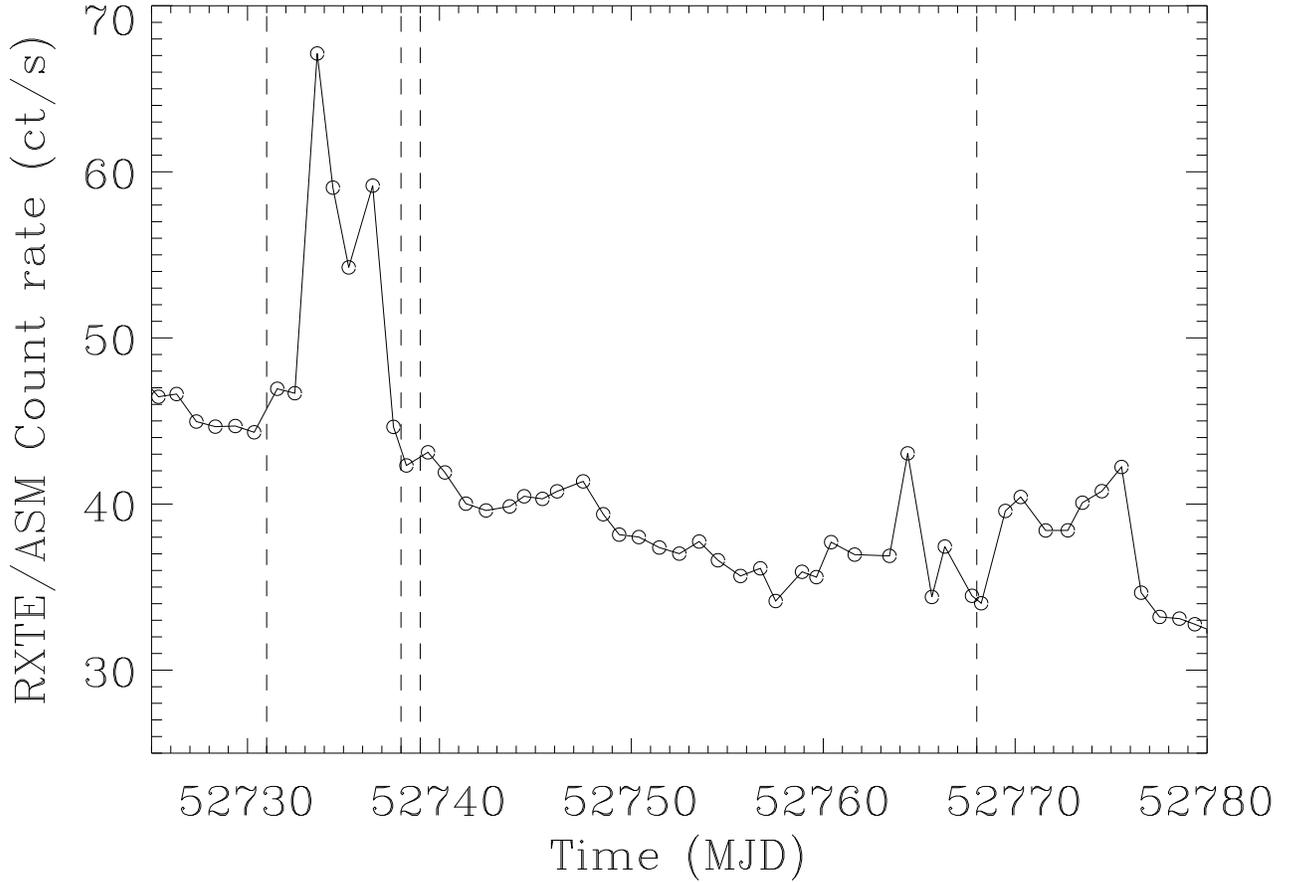}
\caption{RXTE/All Sky Monitor 1.2--12 keV light curve of GRS~1915+105. The vertical dashed lines show
the days our observations took place. The 1.2--12 keV $\sim$1 Crab flare is obvious here. A radio flare
is detected at the same time, with its maximum occurring 2 days later (F03).}
\label{fig:asm}
\end{figure}

\begin{figure}
\plotone{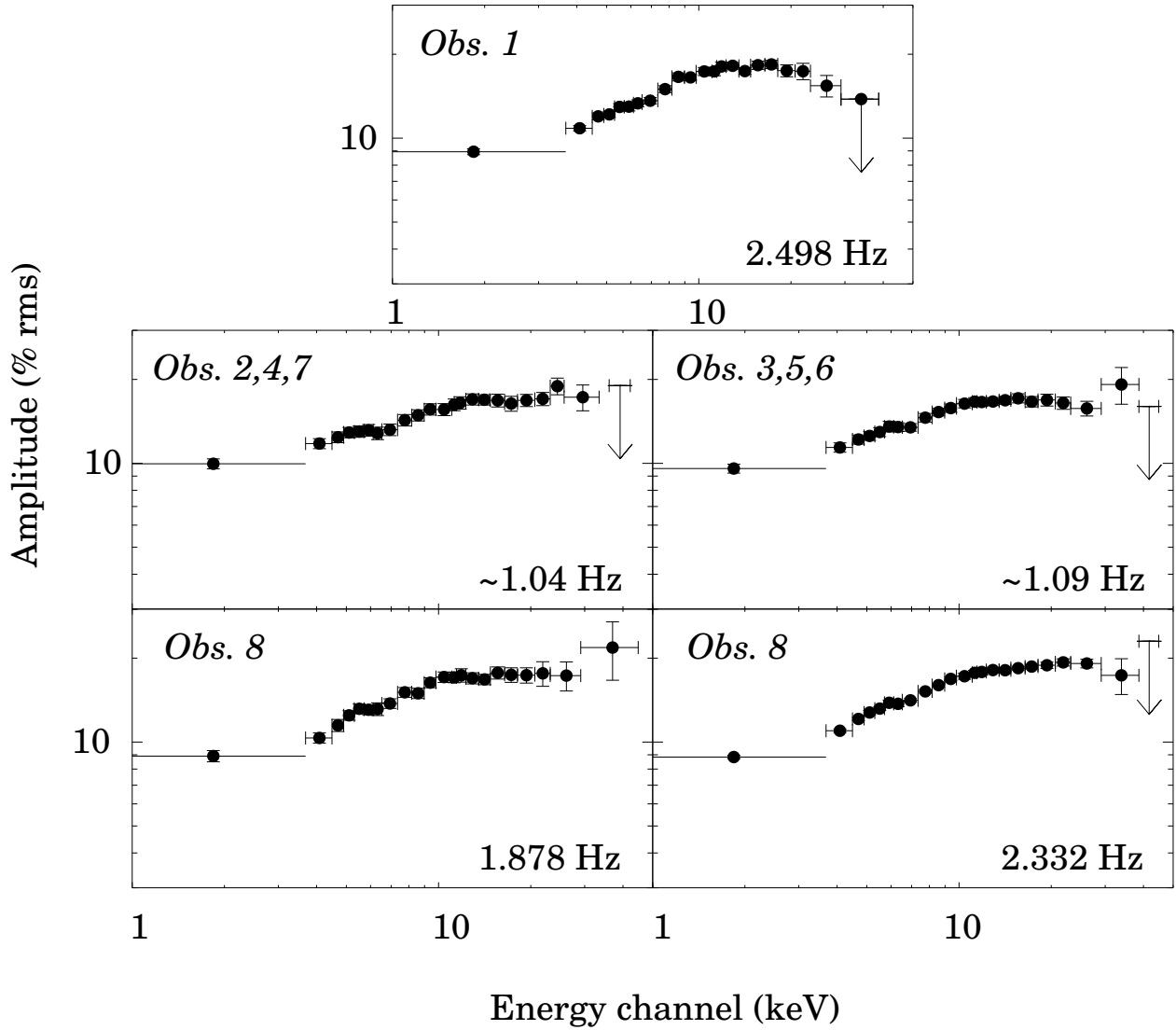}
\caption{Energy dependence of the LFQPO amplitude. The frequency (or mean frequency) of the feature 
and the observation numbers are  written in each panel.}
\label{fig:qpospec}
\end{figure}

The presence of LFQPO in GRS~1915+105 during class $\chi$ (as well as during other
classes) is a known fact (e.g. Muno et al. 1999, Rodriguez et al. 2002a,b).
 It is also known that QPO parameters depend
on spectral parameters in black hole binaries (BHB) in general. 
Here we present observations taken during the same  state, with 
few differences between the spectral parameters returned from the spectral fits.
The parameters of the QPO change dramatically from one observation to another.
Except in Obs. 8, the frequency of the QPO does not seem to correlate to the {\emph{PCA}} 
2--15 keV  flux or  the {\emph{ASM}} 1.2--12 keV flux either (although the highest frequency 
is observed when the {\emph{ASM}} flux is the highest, Fig. \ref{fig:asm}, and Table \ref{tab:qpos}), 
contrary to what is usually claimed/observed. This could indicate some definite
peculiarities in Obs. 1-7 that are taken just before and after the X-ray/radio flare 
(Fig. \ref{fig:asm}, F03). On the other hand, Obs. 8 occurs later,  after GRS~1915+105 
apparently went off the linear decay phase, after the {\emph{ASM}} light curve 
showed some variability again.\\
\indent The most striking behavior appears when studying the energy dependence
of the QPO amplitude. It is expected and a known fact that it presents 
a turn-over at some point  (e.g. Tomsick \& Kaaret 2001, Rodriguez et al. 2002a). 
We report here, for the first time, a clear evolution of the turn-over energy
between states that are spectrally similar, and have similar {\emph{PCA}} fluxes. This ``cut-off'' 
energy has an origin that is unclear. It could represent, for example, some specific
temperature at which the QPO is produced, either through oscillations of a shocked 
boundary layer between the accretion disk, and a hot inner flow (e.g. Chakrabarti 
\& Titarchuk 1995), or by a hot spot orbiting at some specific radius in the disk 
(e.g. Rodriguez et al. 2002a; Tagger et al. 2004). In these two cases, however, 
we would expect the frequency of the QPO to scale with the inner radius 
of the accretion disk, and thus the soft X-ray flux (Molteni, Sponholz \& 
Chakrabarti 1996; Tagger \& Pellat 1999), unless the soft X-rays are not uniquely produced 
by the accretion disk, but by another physical medium, as e.g. a 
compact jet (see Markoff \& Nowak 2004). 
The variations of the soft X-rays flux could be due to variations of the compact jet flux 
(with a steady thermal flux from the accretion disk), as we discuss below. \\
\indent The spectral approach presented in H04, and
Rodriguez et al. (2004), the systematic analysis of type II states 
(Trudolyubov 2001), and the detection of a hard tail up to (at least) 
600 keV with OSSE (Zdziarski et al. 2001) raise the challenging question of the origin of the
third spectral component needed to fit the high energy spectra well. Models 
of jet emission (e.g. Markoff et al. 2003, Markoff \& Nowak 2004)
propose a jet model in which the X-ray spectrum of an XRB would represent the sum of
 thermal emission from the accretion disk, direct synchrotron  from the jet, 
inverse Comptonization (either through synchrotron self-Compton from the jet, or 
Comptonization on the basis of the jet, the ``corona''), reflection of these radiations on 
the accretion disk. It should be noted 
This proposition has found an echo with 
the radio flux/X-ray flux correlation found in several BHBs when in  the
low hard state (when the compact jet is present, e.g. Corbel et al. 2003, Gallo et al. 2003), 
but also in the case of radio loud AGN (e.g. Merloni, Heinz \& di Matteo, 2003; 
Falcke, K\"ording \& Markoff, 2004).
Our RXTE observation of MJD 52731 occurred at a time when the radio flux 
was high and indicative of the presence of the compact jet (F03). 
The level of radio emission as measured by the Ryle telescope at 15~GHz is about
 130-150 mJy during this observation, with a spectrum extending up to the near 
infra-red range (F03). Unfortunately we do not have such a nice 
coverage for the following observations, but the observation of MJDs 52738-52739 
indicates a 
level of 15 GHz emission higher ($\sim 250$~mJy, F03), that is 
dropping rapidly. We remark that this observation occurred just after a radio flare 
indicative of a discrete ejection. It is thus very likely that the radio  emission 
this  day partly originates from the discrete ejection (with a different spectrum). 
During the last observation, the radio 
flux is very low compared to the two previous dates, with a level dropping from 
107 mJy on MJD 52767 to 44 mJy on MJD 52769 (H04). Both our spectral analysis 
(H04, Rodriguez et al. 2004) and the properties of the 
QPOs (present work) can be understood easily, if the  X-ray emission in GRS~1915+105 during 
radio loud/type II/class $\chi_1-\chi_3$ observations originate (as proposed 
by Markoff \& Nowak 2004) from two different physical processes (beside the thermal emission of the 
accretion disk): Comptonization and synchrotron radiation. The high energy spectrum of a
source with a compact jet represents thus the sum of these different emission processes.
As a result the spectrum will therefore strongly depend on the relative contribution of each
of these emission processes.  
The break in the energy spectrum could be representative of the energy at which the relative 
contributions of these components cross each other. Above the break the  
contribution of the synchrotron radiation would be the dominant process to the spectrum.
Then the higher the relative contribution of the 
synchrotron component (to the overall spectrum) the lower the break energy is.
 In this case, to understand the energy dependence
of the QPO amplitude, one has to assume that the QPO is contained in the Comptonized
flux, and not in the synchrotron flux. Then the position of the cut-off in the energy
dependence of the QPO amplitude would be linked to the synchrotron flux emitted by 
the jet. 
We find this interpretation at least qualitatively in good agreement with several 
observational facts:
\begin{itemize}
\item the compact jet model has successfully been used in the fitting of 
different BHBs (e.g. Markoff et al. 2001).
\item type II states show  a 2--30 keV level of variability lower than that of 
type I (radio quiet) states (Trudolyubov 2001). 
\item  a compact jet is detected during the observation showing the clear and well constrained 
cut-off in the energy dependence of the QPO amplitude (Obs. 1, Fig. \ref{fig:qpospec})
\item a high level of radio emission is detected during the observation taken on MJDs 
52738-52739, a turn-over in the energy dependence of the $\sim 1.09$~Hz QPO is detected 
(Fig. \ref{fig:qpospec}, Table \ref{tab:fit}, although it is absent in the spectrum of 
the $\sim1.04$~Hz QPO), while for the last observation (MJD 52768) 
the radio flux is much lower, a single power law can fit the first QPO spectrum, and 
the turn over is poorly constrained (Fig. \ref{fig:qpospec}, Table \ref{tab:fit}).
\end{itemize}
We should add that the Optical/UV/X-ray variability (and presence of LFQPO 
in those bands) seen in XTE J1118+480 (Hynes et al. 2003), a Black Hole X-ray transient in which the compact jet model 
has been shown to fit the broad band spectra well (Markoff et al. 2001), is also 
compatible with our interpretation. Hynes et al. (2003) pointed out that the 
variability could not originate from the disk itself, but involved another non-thermal 
source of photons.\\
\indent The lack of complete simultaneity, between the radio and X-ray observations, 
prevents 
us from drawing any firmer conclusions. In addition, a cut-off in the spectrum of the
compact jet is expected in the near infra-red domain. Knowing its exact position
would allow us to estimate  the flux expected from the jet in the hard 
X-rays accurately, and thus test our hypothesis. We hope to obtain such simultaneous 
coverages in the near future, with {\emph{INTEGRAL}}, {\emph{RXTE}} for the high energies, but also 
the Ryle telescope and the VLA, in the radio domain, and {\emph{Spitzer}}, 
and ground based telescopes in the infra red domain.

\begin{acknowledgements}
The authors would like to thank G. Henri for useful discussions and G. Pooley
for kindly providing the Ryle data to our group. J.R. acknowledges financial support 
from the French Space Agency (CNES). DCH acknowledges the Finnish Academy.
\end{acknowledgements}

\clearpage
\begin{table}
\centering
\caption{Log of the RXTE observations used in the present analysis. The first of these observations was performed simultaneously with the 
multi-wavelength campaign discussed in  Fuchs et al. (2003). Observations are time ordered. $^\star$
cts/s in the top layer of PCU \#2, both anodes.}
\begin{tabular}{cccccc}
\hline
Observation \# & Obs. Date & Obs. Id. & Good Time & Number of & Net Count Rate  \\
               &     (MJD) & (P80127-)&  (s)      &    PCU    & /PCU$^\star$\\
\hline
1 & 52731& 01-03-00 &  9,300 s & 4 & 1737.5 cts/s\\
2 & 52738 & 02-01-00 &  5,400 s & 3 & 1700.8 cts/s\\
3 & 52739 & 02-02-02 &  1,800 s & 3 & 1675.8 cts/s\\ 
4 &              & 02-02-01 &  1,800 s & 4 & 1666.9 cts/s\\
5 &              & 02-02-00 &  2,060 s & 3 & 1660.4 cts/s\\
6 &              & 02-03-00 & 11,100 s & 3-4 & 1674.7 cts/s\\
7 &              & 02-01-01 & 3,200 s & 3 & 1677.0 cts/s\\
8 & 52768 & 03-01-00 & 14,000 & 3-4 & 1460.2 cts/s \\

\hline
\end{tabular}
\label{tab:log}
\end{table}

\begin{table}
\caption{Parameters of the LFQPO detected in each of 
the 8 sequences. $^\star$ 
$Q=\frac{\mathrm{Centroid frequency}}{FWHM}$}
\centering
\begin{tabular}{cccc}
\hline
Obs. sequence & QPO frequency & Q$^\star$ & rms amplitude\\
              &    (Hz)       &           &   \% \\
\hline
1 &2.498$\pm 0.005$  & 5.0 & 12.6$\pm0.3$ \\
2 &1.040$\pm0.004$  & 3.8 &  13.5 $\pm0.6$\\
3 & 1.081$\pm0.004$ & 7.2 &  13.2$^{+1.0}_{-0.9}$\\
4 & 1.039$\pm0.004$ & 7.2  & 12.7 $^{+1.2}_{-1.3}$\\
5 & 1.097$\pm0.005$ & 10.0 & 11.5$\pm1.0$ \\
6 & 1.096$\pm0.002$ & 6.4 & 13.4$\pm0.4$ \\
7 &1.060$\pm0.003$  & 6.4  & 13.0$\pm 0.9$\\
8\_QPO1 & 1.878$_{-0.003}^{+0.005}$ & 6.8 & 12.9$\pm0.5$ \\
8\_QPO2 &2.332$\pm0.005$  & 5.3 &  14.9$\pm0.5$\\
\hline
\end{tabular}
\label{tab:qpos}
\end{table}

\begin{table}
\caption{Parameters of the fits to the energy dependence of the 
QPO amplitude, also refer to as QPO spectra in the text. The errors are given at the 
one-$\sigma$ level. $^\star$ No cut-off is detected in this observation at least up to $\sim40$ keV.}
\centering
\begin{tabular}{cccc}
\hline
QPO freq. & Power-law  slope & Cut-off energy & reduced $\chi^2$ (dof)\\
  (Hz)    &    ($\Gamma$)     &  (keV) &  \\
\hline
2.498 & $-0.77\pm0.04$  & $21.9_{-2.2}^{+2.7}$ & 1.0 (19) \\
1.04 & $-0.26\pm 0.02$  & No cut-off$^\star$ & 0.92 (21)\\
1.09 & $-0.59\pm 0.05$  & $29.5_{-3.9}^{+5.2}$ & 0.37 (20)\\
1.878 & $-0.70\pm0.07$ & $25.6_{-4.4}^{+6.5}$ & 0.5 (19) \\
2.332 & $-0.71\pm0.05$  & $26.7_{-3.3}^{+4.4}$ & 0.28 (20)\\
\hline
\end{tabular}
\label{tab:fit}
\end{table}

\clearpage

\end{document}